\documentclass[aps,prl,twocolumn,superscriptaddress]{revtex4}  

\usepackage{amsmath}
\usepackage{amssymb}
\usepackage{graphicx}
\usepackage{epstopdf}
\usepackage{color}
\usepackage{cancel}

\include{papersdefinitions}

\def\bra#1{\left<#1\right|} 
\def\ket#1{\left|#1\right>}
\newcommand{\be}{\begin{equation}}
\newcommand{\ee}{\end{equation}}
\newcommand{\bea}{\begin{eqnarray}}
\newcommand{\eea}{\end{eqnarray}}

\begin{document}

\title{Single-photon-level quantum memory at room temperature}

\author{K. F. Reim}
\affiliation{Clarendon Laboratory, University of Oxford, Parks Road, Oxford OX1 3PU, UK}

\author{P. Michelberger}
\affiliation{Clarendon Laboratory, University of Oxford, Parks Road, Oxford OX1 3PU, UK}

\author{K.C. Lee}
\affiliation{Clarendon Laboratory, University of Oxford, Parks Road, Oxford OX1 3PU, UK}

\author{J. Nunn}
\affiliation{Clarendon Laboratory, University of Oxford, Parks Road, Oxford OX1 3PU, UK}

\author{N. K. Langford}
\affiliation{Clarendon Laboratory, University of Oxford, Parks Road, Oxford OX1 3PU, UK}

\author{I. A. Walmsley}
\email[]{i.walmsley1@physics.ox.ac.uk}
\affiliation{Clarendon Laboratory, University of Oxford, Parks Road, Oxford OX1 3PU, UK}

\date{\today}

\begin{abstract}
Quantum memories capable of storing single photons are essential building blocks for quantum information processing, enabling the storage and transfer of quantum information over long distances~\cite{Duan_2001_N_Long-distance-quantum-communic,Briegel_1998_PRL_Quantum-Repeaters:-The-Role-of}.  
Devices operating at room temperature can be deployed on a large scale and integrated into existing photonic networks, but so far warm quantum memories have been susceptible to noise at the single photon level ~\cite{Manz_2007_PR_Collisional-decoherence-during}. This problem is circumvented in cold atomic ensembles~\cite{Choi_2008_N_Mapping-photonic-entanglement-,Simon_2007_PRL_Interfacing-Collective-Atomic-,Matsukevich_2006_PRL_Deterministic-Single-Photons-v}, but these are bulky and technically complex.
Here we demonstrate controllable, broadband and efficient storage and retrieval of weak coherent light pulses at the single-photon level in warm atomic caesium vapour using the far off-resonant Raman memory scheme~\cite{Reim_2010_P_Towards-high-speed-optical-qua}. The unconditional noise floor is found to be low enough to operate the memory in the quantum regime at room temperature.

\end{abstract}

\maketitle

In an age of ever-increasing, global information transfer, there is growing demand for secure communication technology, such as could be provided by photonic quantum communications networks \cite{Gisin_2002_RMP_Quantum-cryptography}.  Currently, the biggest challenge for such networks is distance. Over short distances, photons, interacting only weakly with their environment, easily and reliably carry quantum information without much decoherence, but intercontinental quantum communication will require quantum repeaters embedded in potentially isolated locations, because photon loss rises otherwise exponentially with distance~\cite{Duan_2001_N_Long-distance-quantum-communic,Briegel_1998_PRL_Quantum-Repeaters:-The-Role-of}.  In general, these repeaters will require some sort of quantum memory, a coherent device where single photons are reversibly coupled into and out of an atomic system, to be stored, possibly processed and then redistributed.  In order to be practically useful, this will need to have sufficiently large bandwidth, high efficiency and long storage time, with multimode capacity, and a low-enough noise level to enable operation at the quantum level.  Furthermore, to be truly practical, a critical feature of such memories will be ease of operation near room temperature for use in isolated, potentially unmanned locations. 

Warm atomic vapours of alkali atoms such as rubidium and caesium are potentially excellent storage media. At moderate temperatures the vapour pressure is sufficient to render the doublet lines optically thick, providing a strong atom-photon interaction in a centimeter-scale vapour cell. With the addition of a buffer gas to slow atomic diffusion, collective atomic coherences on the order of milliseconds have been observed \cite{Julsgaard:2004lr}. There is no need for laser trapping, high vacuum or cryogenic cooling, because atomic vapour cells can be miniaturised \cite{Baluktsian:2010kx}, integrated onto chips using hollow-core waveguide structures \cite{Wu:2010uq} and mass-produced. Cost-effective integration is likely to be crucial for quantum photonics to develop into a mature technology.

A number of experiments using electromagnetically induced transparency (EIT) have demonstrated efficient storage and recall of optical signals in warm atomic vapour \cite{Novikova:2007kc}, and conditional storage and recall of heralded single photons \cite{Eisaman_2005_N_Electromagnetically-induced-tr}. In EIT, however, the frequency of the signal matches the atomic resonance, and collision-induced fluorescence at this frequency makes the unconditional (i.e.\ unheralded) noise floor of this protocol too high for quantum applications~\cite{Manz_2007_PR_Collisional-decoherence-during}.

These issues are avoided in cold-atom experiments~\cite{Chou_2007_S_Functional-Quantum-Nodes-for-E, Chen_2008_P_Memory-built-in-quantum-telepo, Chen_2006_PRL_Deterministic-and-Storable-Sin}, where atomic collisions are suppressed or eliminated. But the technical complexity of these experiments, which require the trapping and cooling of atomic clouds under high vacuum, makes large-scale deployment of this type of memory rather challenging, especially outside controlled laboratory conditions. Similarly, solid-state memories are promising candidates for efficient and low-noise photon storage \cite{Hedges_2010_N_Efficient-quantum-memory-for-l,Riedmatten:2008ek}, but currently they must be operated at cryogenic temperatures.

By contrast, ensemble-type Raman memories which are tuned far from resonance~\cite{Nunn_2007_PRAMOP_Mapping-broadband-single-photo} provide a possible path to fulfilling all of the requirements for quantum-ready operation at room temperature. The far off-resonant Raman interaction results in: (i) extremely broadband capability, allowing to interface the memory with conventional parametric downconversion sources  \cite{Cohen_2009_PRL_Tailored-Photon-Pair-Generatio}; (ii) the ability to optically switch the memory in and out of the quantum channel, or alternatively set the storage level to 50\%, providing a straightforward noninterferometric path to creating light-atom entanglement; (iii) very weak fluorescence noise which is predominantly non-synchronous with the short signal pulse; and (iv) memory efficiency which is insensitive to inhomogeneous broadening, allowing room-temperature operation and a path towards integrated implementations. Apart from the advantage of simplicity~\cite{Raab_1987_PRL_Trapping-of-Neutral-Sodium-Ato}, operating at room temperature also makes it easier to achieve larger optical depths, and hence higher efficiencies~\cite{Hosseini:2009lq,Nunn_2007_PRAMOP_Mapping-broadband-single-photo}.

Recently, we implemented a Raman memory in a hot caesium-vapour atomic ensemble, demonstrating extremely broadband, coherent operation under room-temperature conditions~\cite{Reim_2010_P_Towards-high-speed-optical-qua}. In this letter, we address the remaining requirements for a memory to be able to function usefully in a genuine quantum application. Specifically, we demonstrate total memory efficiencies of $>30\%$, and with only moderate magnetic shielding, we show storage times of up to 4 $\mu$s, around 2,500 times longer than the duration of the pulses themselves. This is already sufficient to improve heralded multiphoton rates from parametric down-conversion sources \cite{Bouwmeester_Experimental_quantum_teleportation}. Finally, we make a detailed analysis of the unconditional noise floor of the memory, which is found to be $<0.25$ photons per pulse; that is --- low enough for quantum applications. The combination of these results shows that we have a quantum-ready memory, capable of handling quantum information in a simple room-temperature design.
  
\begin{figure}
\begin{center}
\includegraphics[scale=0.3]{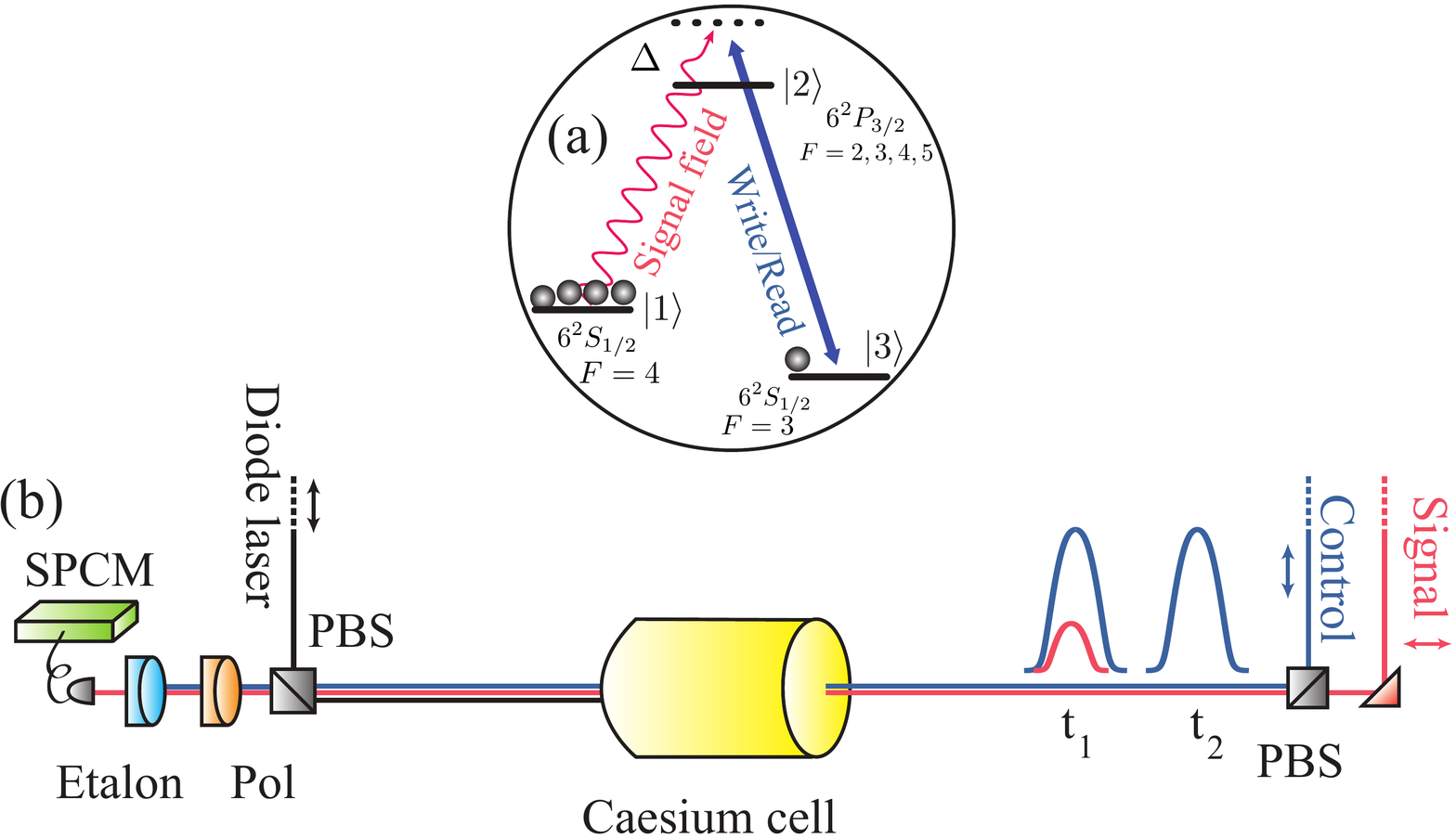}
\caption{(a) $\Lambda$-level scheme for Raman memory. $\Delta$ is the detuning from the atomic resonance. (b) Experimental setup. At $t_1$ the single photon level signal is mapped by a strong write pulse into a spin-wave excitation in the caesium vapour cell. At $t_2$ a strong read pulse reconverts the excitation into a photonic mode. After polarization filtering (Pol) and spectral filtering with Fabry-Perot etalons, the retrieved signal is detected with single Photon counting modules (SPCM). Vertical polarization is indicated as ($\updownarrow$) and  horizontal polarization as ($\leftrightarrow$).}
\label{fig:experiment}
\end{center}
\end{figure}
 
The heart of the quantum memory is a caesium-vapour atomic ensemble, prepared in a vapour cell heated to 62.5$^\circ$C, that makes use of the 852-nm $D_2$ line, with the $6^2 S_{1/2}$ hyperfine states serving as the ground $\ket{1}$ and storage $\ket{3}$ states (Fig. \ref{fig:experiment}(a)). 
At $t_1$, the signal gets mapped by the strong write field into a collective atomic spinwave excitation, and at $t_2$, a strong read pulse reconverts the atomic coherence back into the photonic mode (Fig.~\ref{fig:experiment}(b)).  The signal is separated from the strong write and read fields via spectral and polarisation filtering, which is particularly important when operating at the single-photon level, and then detected using a silicon avalanche photodiode (see Methods for details).  

\begin{figure}
\begin{center}
\includegraphics[width=1\columnwidth]{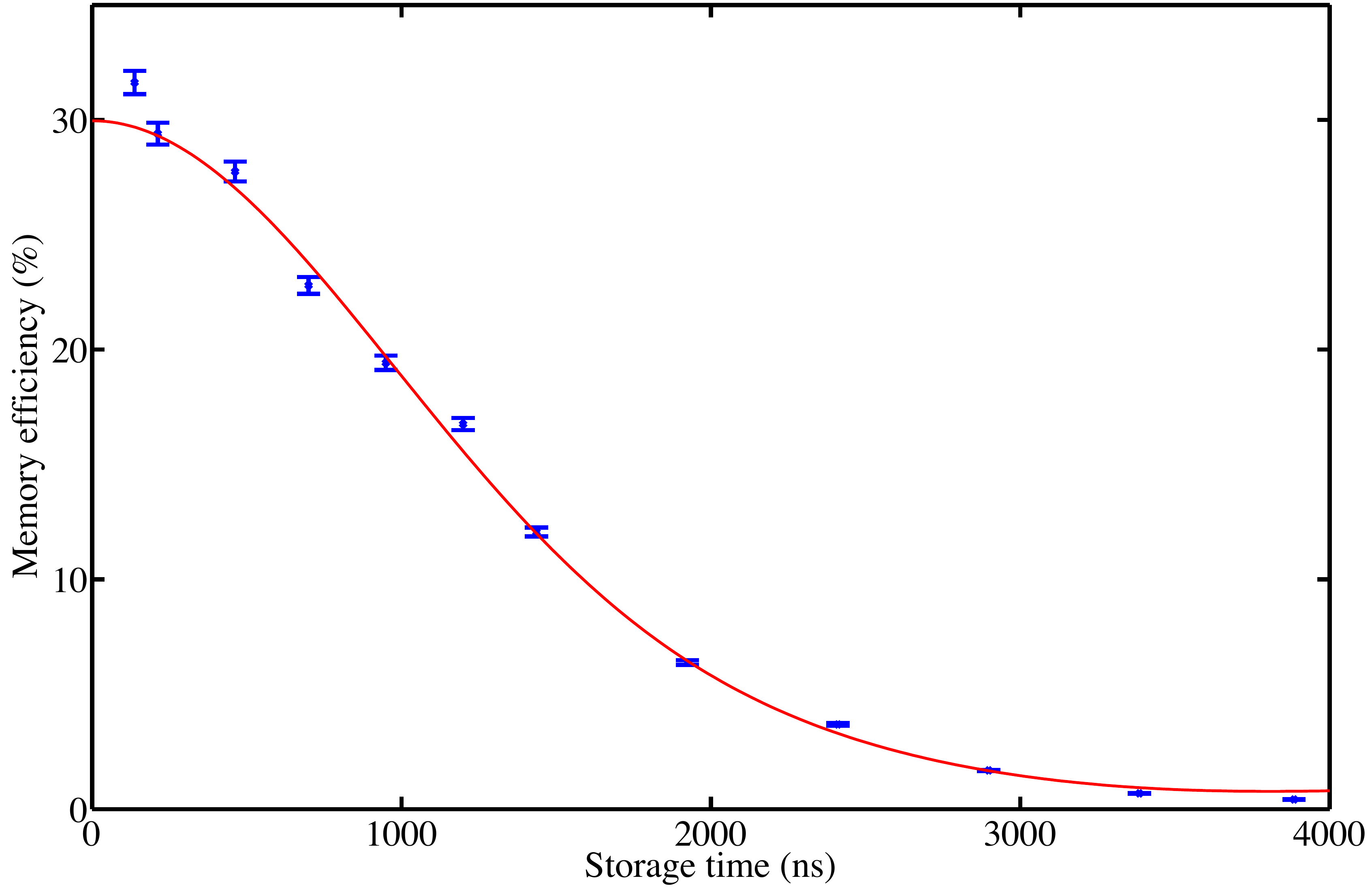}
\caption{Memory efficiency. The lifetime of the memory is about 1.5~$\mu$s. The blue dots indicate experimental data measured with a fast APD. Error bars represent the standard error in the mean. The solid line is the theoretical dephasing predicted for a constant magnetic field of $0.13\pm 0.05$ Gauss, which can be attributed to the residual of the Earth's magnetic field (Supplementary Information). Note that this dephasing could be compensated using spin-echo techniques, or suppressed with improved magnetic shielding.}
\label{fig:storage}
\end{center}
\end{figure}

In the current experiment, we demonstrate total efficiencies around $30\%$ and a memory lifetime around 1.5~$\mu$s (Fig.~\ref{fig:storage}), which is double the efficiency and more than two orders of magnitude improvement in lifetime over the values obtained in our previous experiment~\cite{Reim_2010_P_Towards-high-speed-optical-qua}. These results were obtained with only moderate magnetic shielding and residual static fields are the main dephasing mechanism (see Supplementary Information). In principle, magnetic dephasing can be eliminated by improved shielding, enabling storage times limited by atomic diffusion to several hundred $\mu$s~\cite{Camacho_2009_P_Four-wave-mixing-stopped-light}. 
However, because this memory has such a broad bandwidth, this measured lifetime already corresponds to a time-bandwidth product of $\sim$2500.  Indeed, with memory efficiencies of $20\%$ at 1~$\mu$s, this memory could already be used to improve heralded photon-pair rates with typical parametric down-conversion sources with heralded single photon rates of $\sim$1~MHz.
As discussed in Ref.~\cite{Reim_2010_P_Towards-high-speed-optical-qua}, the memory efficiency is restricted mainly by control field  power and the less efficient, but experimentally simpler, forward-retrieval configuration \cite{Nunn_2007_PRAMOP_Mapping-broadband-single-photo}.

\begin{figure}
\begin{center}
\includegraphics[scale=0.25]{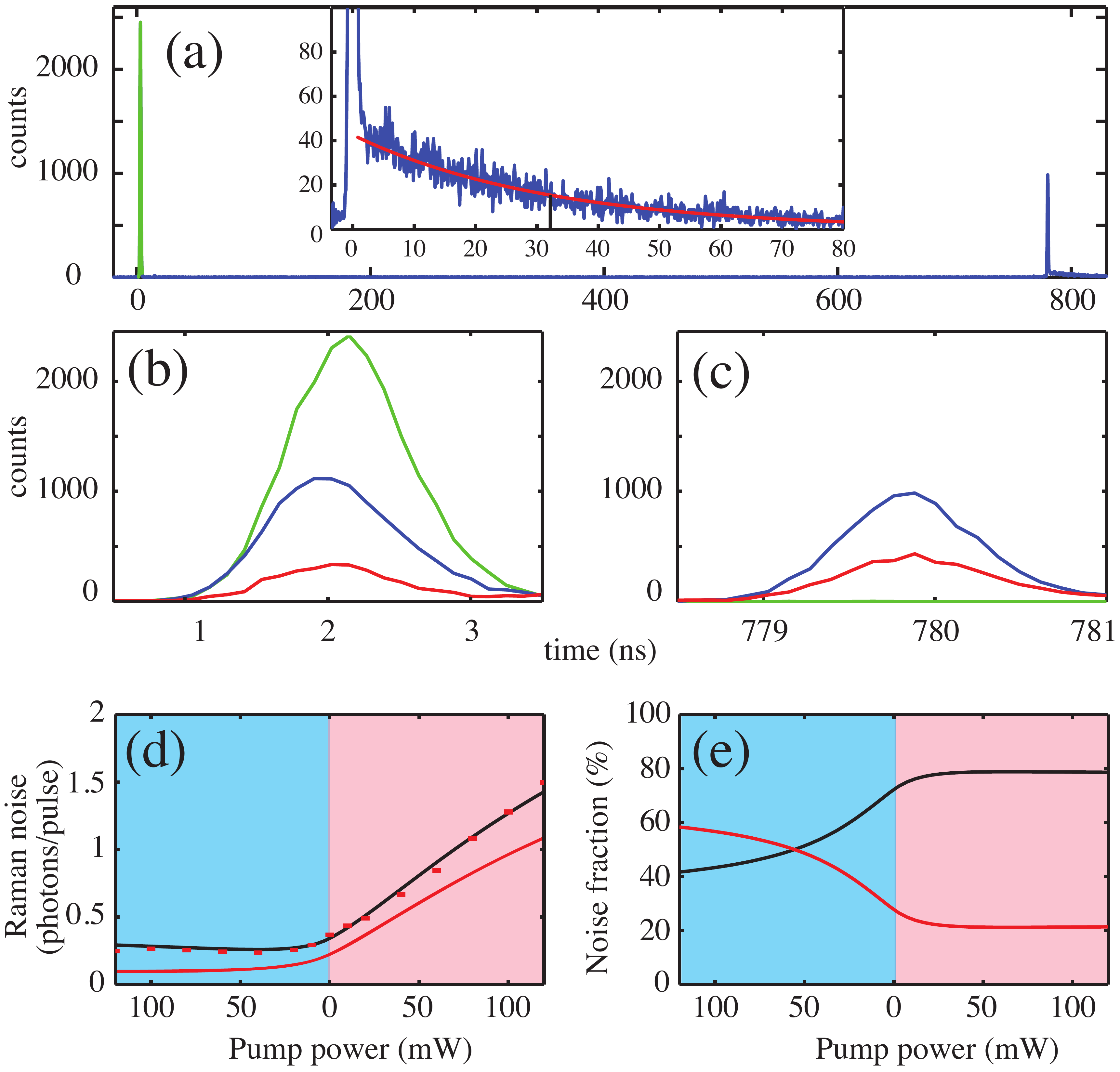}
\caption{Single-photon-level data. (a) Incident ($t_1\,{\sim}\,0$~ns) and retrieved ($t_2\,{\sim}\,780$~ns) pulse. The inset is a zoom of the retrieved pulse showing a fluorescence noise tail. (b,c) Zooms around $t_1$ and $t_2$, showing the: incident signal (transmitted/retrieved signal with no control field) (green); transmitted/retrieved signal with the control field (blue); and noise (control field only) (red). Histograms are accumulated over 360,000 runs. (d) Blue shaded area: Optical pumping on the $\ket{3}\leftrightarrow\ket{2}$ (blue) transition. Red shaded area:  Optical pumping on the $\ket{1}{\leftrightarrow}\ket{2}$ (red) transition. The red points are data and the black line is a theoretical prediction of noise due to spontaneous Raman scattering. The red solid line shows the predicted noise due only to Stokes scattering (e) Predicted fractional noise contributions from Stokes (black curve) and anti-Stokes (red curve) light. Theoretical details can be found in the supplementary information.}
\label{fig:single_photon}
\end{center}
\end{figure}

Next, we measure the memory output using a single-photon counting module (SPCM) to test its performance at the single-photon level. Figure \ref{fig:single_photon}(a) shows storage at $t=0$~ns and retrieval at $t=780$~ns later for an average input signal of 1.6 photons per pulse ($< 1$ retrieved photon per pulse), while Figs~\ref{fig:single_photon}(b,c) show the storage and retrieval processes in detail. The strong reduction in transmitted signal (Fig.~\ref{fig:single_photon}(b)) between the control field being off (green) and on (blue), and the significant amount of retrieved signal (Fig.~\ref{fig:single_photon}(c), blue) demonstrates that the quantum memory operates well at the single-photon level.  However, to properly characterise the noise in these signals, it is critical to measure the \emph{unconditional noise floor}: the signal detected when memory retrieval is triggered without any stored input signal (Figs~\ref{fig:single_photon}(b,c), red curves). In our case, this is the detected signal when the control field is sent in with no input signal, and is measured to be 0.25 photons per pulse. Already substantially less than a single photon, this enables operation in the quantum regime, even with $30\%$ memory efficiency.

We now explore the origin of the noise photons to determine if the observed noise could be further reduced. The inset to Fig.~\ref{fig:single_photon}(a) shows that the retrieved signal peak is followed by an exponential tail, which is fluorescence noise from the excited state, $\ket{2}$: this process is excited by the control field, even far from resonance, in the presence of atomic collisions. The red curve is a least-squares fit to the tail of the experimental data and yields an estimated lifetime of 32$\pm$2~ns (expected value 30.5~ns  \cite{Steck_2009_AOH_Cesium-D-Line-Data}; error bars derived from multiple experiments). While such collision-induced fluorescence limits the usefulness of other more narrowband room-temperature memories (as shown in \cite{Manz_2007_PR_Collisional-decoherence-during}), in our extremely broadband memory, the 30 ns time scale of these emissions is much longer than the duration of the readout event (set by the 300 ps pulse width and detector timing jitter). By time gating the detection, fluorescence therefore made a negligible contribution to the measured, instantaneous noise floor (Figs~\ref{fig:single_photon}(b,c)).

To understand where the instantaneous noise comes from, we then investigated its dependence on optical pumping. The blue shading in Fig.~\ref{fig:single_photon}(d) indicates optical pumping on the $\ket{3}{\leftrightarrow}\ket{2}$ (``blue'') transition, whereas the red shading represents optical pumping on the $\ket{1}{\leftrightarrow}\ket{2}$ (``red'') transition. Increasing the pump power on the blue transition partially suppresses the noise, although it rapidly levels off at an average of 0.25 photons per pulse, while the noise level rises linearly with increasing pump power on the red transition.

These observations are well-described by a simple noise model (supplementary information) based on \emph{spontaneous Raman scattering} (SRS) \cite{Lukin:1999uq,Phillipsa:2009fk}, in which Stokes and anti-Stokes photons scattered during the control pulse are both transmitted through our etalon-based filters (see Methods section) and detected as noise. Since the model does not include collision-induced fluorescence \cite{Manz_2007_PR_Collisional-decoherence-during} or leakage of the control, this suggests that these contributions are negligible. Furthermore, as shown in Fig.~\ref{fig:single_photon}(d) and (e), around 60\% of the noise affecting the quantum memory is emitted at the anti-Stokes frequency (when operated with maximal blue pumping) and could be removed using further spectral filtering. This would already bring the true unconditional noise floor down to 0.1 photons per pulse resulting in an unconditional signal to noise ratio (SNR) of 10:1 for single photon retrieval. The remaining signal-frequency (Stokes) noise remains even if the optical pumping and spectral filtering are perfect and has its origin in four-wave mixing seeded by spontaneous anti-Stokes scattering. However, even this noise can be eliminated if the anti-Stokes channel can be suppressed or rendered much weaker than the Stokes channel, for instance by operating the memory closer to resonance, so that the anti-Stokes detuning is relatively much larger. Finally, although our fibre-coupled detection system is optimised for the mode of the signal, the noise is scattered mostly into the control mode \cite{Raymer:1985qy,Sorensen:2009fj}, which suggests that the noise floor can be further reduced by angle tuning the control field \cite{Surmacz:2008vf}. These results show that the Raman memory, therefore, represents a genuine quantum-ready optical memory, functioning at room temperature. 

In conclusion, we have demonstrated a coherent broadband single photon level optical memory at room temperature with a detailed investigation of the unconditional single-photon noise floor. We have also extended the memory storage time far enough that it could be used to improve heralded multiphoton rates from parametric down-conversion sources --- an important step towards making this type of memory useful for quantum repeaters. This work shows that the far off-resonant Raman memory scheme makes it possible to implement a quantum ready memory in warm atomic vapour, and together with technological advances in microfabrication of vapour cells and hollow-core waveguide structures embedded on chips, opens a path to an integrated, scalable quantum information technology architecture. 

\section*{Methods}

The atomic ensemble is initially prepared in the ground state~$\ket{1}$ by optically pumping with an external cavity diode laser (ECDL). For the storage process, the signal and write pulses, both derived from a Ti:Sapph laser oscillator and an EOM \cite{Reim_2010_P_Towards-high-speed-optical-qua}, are sent temporally and spatially overlapped into the  vapour cell (Fig. \ref{fig:experiment}(b)). The pulses have a duration of 300 ps corresponding to a bandwidth of 1.5~GHz. The bandwidth of the memory is defined dynamically by the strong write field which is about $10^9$ times brighter than the single photon level signal field. 
The necessary filtering is provided by a Glan-laser polariser and three air-spaced Fabry-Perot etalons with a free spectral range of 18.4~GHz and transmission window of 1.5~GHz, giving a total extinction ratio of $10^{-11}$. Note that the free spectral range is twice the 9.2~GHz Stokes shift, so both Stokes (signal) and anti-Stokes (noise) frequencies pass through the filters. As discussed in the main text, anti-Stokes noise could be eliminated using more selective filters. The detection system consists of single photon counting modules (SPCM) combined with a time-to-amplitude converter (TAC) and a multichannel analyser (MCA), which allows photo-detection events to be time-binned with a resolution of $\sim125$~ps. All experimental single photon level data are histograms derived from 360,000 experiments.  The memory operates at a repetition rate of 3 kHz and the storage time is set to 800~ns. 

\section*{Acknowledgements}
This work was supported by the EPSRC through the QIP IRC (GR/S82716/01) and project EP/C51933/01. KFR was supported by the Marie-Curie-Network EMALI. PM was supported by FASTQUAST. IAW was supported in part by the European Commission under the Integrated Project Qubit Applications (QAP) funded by the IST directorate as Contract Number
015848, and the Royal Society.

\section*{Author contribution}
KFR and PM carried out the experimental work and took the data. KFR, JN and NKL were responsible for the experimental design and theoretical analysis, and KCL and IAW contributed to the experiment. The manuscript was written by KFR with input from JN, NKL and IAW.

\section*{Supplementary information}

\subsection*{SRS noise model}
The dominant noise process in our Raman memory appears to be spontaneous Stokes and anti-Stokes scattering during the strong control pulse. To model the noise, we consider the Maxwell-Bloch equations describing the collinear propagation of Stokes and anti-Stokes field amplitudes $A_\mathrm{S}$, $A_\mathrm{AS}$ through a $\Lambda$-type ensemble with resonant optical depth $d$ along the $z$-axis, in the presence of a control pulse, whose shape is described by the time dependent Rabi frequency $\Omega = \Omega(\tau)$, where $\tau$ is the time in a reference frame moving with the control. In the adiabatic limit that the bandwidth and Rabi frequency of the control are much smaller than the detuning of the fields from resonance, the propagation equations take the form
\begin{eqnarray}
\label{MaxBloch}  \left[\partial_z + \frac{d\gamma p_1}{\Gamma_\mathrm{S}}\right]A_\mathrm{S} &=& -\frac{\Omega \sqrt{d\gamma}}{\Gamma_\mathrm{S}}B,\\
\nonumber \left[\partial_z -\frac{d\gamma p_3}{\Gamma_\mathrm{AS}^*}\right]A^\dagger_\mathrm{AS} &=& -\frac{\Omega^*\sqrt{d\gamma}}{\Gamma_\mathrm{AS}^*}B,\\
\nonumber \left[\partial_\tau - |\Omega|^2\left(\frac{1}{\Gamma_\mathrm{S}}-\frac{1}{\Gamma_\mathrm{AS}^*}\right)\right]B&=&-\sqrt{d\gamma}\Omega^*\left(\frac{p_1}{\Gamma_\mathrm{S}} + \frac{p_3}{\Gamma_\mathrm{S}^*}\right)A_\mathrm{S}\\
\nonumber & &-\sqrt{d\gamma}\Omega\left(\frac{p_1}{\Gamma_\mathrm{AS}} + \frac{p_3}{\Gamma_\mathrm{AS}^*}\right)A_\mathrm{AS}^\dagger,
\end{eqnarray}
where $B$ is the amplitude of the spin wave, $\Gamma_{\mathrm{S},\mathrm{AS}} = \gamma-\mathrm{i} \Delta_{\mathrm{S},\mathrm{AS} }$ is the complex detuning of the Stokes (anti-Stokes) field, $\gamma$ is the homogeneous linewidth of the $\ket{1}{\leftrightarrow}\ket{2}$ transition, $\Delta_\mathrm{AS} = \Delta_\mathrm{S}+\delta$, the Stokes detuning $\Delta_\mathrm{S}$ is equal to the detuning $\Delta$ of the signal field in Fig.~\ref{fig:experiment}, and $\delta = 9.2$~GHz is the Stokes shift. The $z$-coordinate has been normalised so that it runs from $0$ to $1$. We have defined $p_1$ ($p_3$) as the fraction of atoms initially in state $\ket{1}$ ($\ket{3}$). Since the noise process is weak, we assume that these populations, as well as the control pulse, are unaffected by the interaction.

In general, the solution is given by
\begin{eqnarray}
\nonumber A_{\mathrm{S},\mathrm{out}}(\tau) &=& \int_{-\infty}^\infty K_\mathrm{S}(\tau,\tau')A_{\mathrm{S},\mathrm{in}}(\tau')\,\mathrm{d}\tau' \\
\nonumber && + \int_{-\infty}^\infty G_\mathrm{S}(\tau,\tau')A_{\mathrm{AS},\mathrm{in}}^\dagger(\tau')\,\mathrm{d}\tau' \\
\nonumber && + \int_0^1L_\mathrm{S}(\tau,z)B_\mathrm{in}(z)\,\mathrm{d}z,\\
\nonumber A_{\mathrm{AS},\mathrm{out}}(\tau) &=& \int_{-\infty}^\infty K_\mathrm{AS}(\tau,\tau')A_{\mathrm{AS},\mathrm{in}}(\tau')\,\mathrm{d}\tau' \\
\nonumber && + \int_{-\infty}^\infty G_\mathrm{AS}(\tau,\tau')A_{\mathrm{S},\mathrm{in}}^\dagger(\tau')\,\mathrm{d}\tau' \\
\nonumber && + \int_0^1L_\mathrm{AS}(\tau,z)B_\mathrm{in}^\dagger(z)\,\mathrm{d}z,
\end{eqnarray}
where the subscripts `$\mathrm{in}$' (`$\mathrm{out}$') describe the amplitudes at the start (end) of the interaction, and where the integral kernels $K_{\mathrm{S},\mathrm{AS}}$, $G_{\mathrm{S},\mathrm{AS}}$ and $L_{\mathrm{S},\mathrm{AS}}$ are Green's functions that propagate the input fields to compute the output fields. Both the Stokes and anti-Stokes frequencies are passed by our spectral filters, so our noise signal is calculated by adding the average number of photons scattered into both Stokes and anti-Stokes modes,
\begin{eqnarray}
\nonumber
S &=& \int_{-\infty}^\infty \langle A_{\mathrm{S},\mathrm{out}}^\dagger(\tau)A_{\mathrm{S},\mathrm{out}}(\tau)\rangle\,\mathrm{d}\tau\\
\label{noise} && + \int_{-\infty}^\infty \langle A_{\mathrm{AS},\mathrm{out}}^\dagger(\tau)A_{\mathrm{AS},\mathrm{out}}(\tau)\rangle \,\mathrm{d}\tau.
\end{eqnarray}
It can be shown that the averaged photon numbers do not depend on the shape of the control pulse, but only on its energy, through the quantity $W = \int_{-\infty}^\infty |\Omega(\tau)|^2\,\mathrm{d}\tau$ \cite{Nunn_2007_PRAMOP_Mapping-broadband-single-photo}, as is usual in the transient regime of spontaneous Raman scattering \cite{Raymer:1985qy}. The initial state used to evaluate the expectation values is the vacuum state, with no Stokes/anti-Stokes photons and no spin wave excitations. Using the bosonic commutation relations $[A_{\mathrm{A},\mathrm{AS},\mathrm{in}}(\tau),A_{\mathrm{A},\mathrm{AS},\mathrm{in}}^\dagger(\tau')]=\delta(\tau-\tau')$, and noting that $B\propto \ket{1}\!\bra{3}$, so that $\langle B_\mathrm{in}^\dagger(z) B_\mathrm{in}(z')\rangle = p_3\delta(z-z')$ and $\langle B_\mathrm{in}(z) B_\mathrm{in}^\dagger(z')\rangle = p_1\delta(z-z')$, we obtain
\begin{eqnarray}
\label{Sexp} S &=& \int_{-\infty}^\infty \int_{-\infty}^\infty \mathrm{d}\tau' \mathrm{d}\tau \left\{|G_\mathrm{S}(\tau,\tau')|^2+|G_\mathrm{AS}(\tau,\tau')|^2\right\}\\
\nonumber && +  \int_{-\infty}^\infty \int_0^1 \mathrm{d}z \mathrm{d}\tau \left\{p_3|L_\mathrm{S}(\tau,z)|^2+p_1|L_\mathrm{AS}(\tau,z)|^2\right\}.
\end{eqnarray}
This expression depends only on the Green's functions, and can be evaluated by solving the system Eqs.~(\ref{MaxBloch}) numerically, treating the amplitudes $A_\mathrm{S}$, $A_\mathrm{AS}$, and $B$ as classical complex-valued functions. The variation of the populations $p_{1,3}$ with optical pumping power $P$ is modelled by setting
\begin{equation}
\label{pump}
p_3(P) = \frac{1}{2}\left[1+\frac{P/P_\mathrm{s}}{1+|P|/P_\mathrm{s}}\right];\quad p_1(P) = 1-p_3(P),
\end{equation}
where $P_\mathrm{s}$ is the saturation power and with the convention that negative values of $P$ indicate pumping on the blue $\ket{2}\leftrightarrow\ket{3}$ transition. The observed spontaneously scattered Raman noise is then given by $S_{\rm obs}(P)=\kappa S(P)$, which incorporates a scaling parameter into the simple one-dimensional theory to account for the fact that the Raman scattering process has a broad spatial distribution \cite{Raymer:1985qy,Sorensen:2009fj}, whereas the signal is emitted into a single narrower spatial mode. The parameter $\kappa$ defines the overlap between the scattered mode and the detected mode, which is defined and filtered by a single-mode fibre optimised on the signal mode. For our experiment we have $d=1900$, $\gamma = 16$~MHz, $\Delta=15$~GHz and $W=30$~GHz. To correctly describe our noise measurements, we find $P_\mathrm{s}=84$ mW and $\kappa = 0.12$ using a least-squares fit.

\subsection*{Magnetic dephasing}
We model the magnetic dephasing by considering the evolution of the Raman coherence excited in the memory under the influence of a static magnetic field. Suppose that an atom is initially prepared in the Zeeman state $\ket{F_\mathrm{i},m_\mathrm{i}}$ with probability $p_{m_\mathrm{i}}$ within the initial hyperfine manifold with $F_\mathrm{i}=4$. Storing a signal pulse creates a spin wave, represented by applying the operator $S^\dagger$ to the initial atomic state, where $S = \alpha I + \beta \Sigma$, with $I$ the identity operator, $\alpha$, $\beta$ two coefficients quantifying the amplitude of the spin wave, whose values are not important, and $\Sigma$ the transition operator given by
\begin{equation}
\label{Sigma}
\Sigma = \sum_{m_\mathrm{i}=-F_\mathrm{i}}^{F_\mathrm{i}} \sum_{m_\mathrm{f}=-F_\mathrm{f}}^{F_\mathrm{f}} C(m_\mathrm{i},m_\mathrm{f})\ket{F_\mathrm{i},m_\mathrm{i}}\!\!\bra{F_\mathrm{f},m_\mathrm{f}}.
\end{equation}
Here $C(m_\mathrm{i},m_\mathrm{f})$ is the coupling coefficient between the initial state and the final Zeeman state $\ket{F_\mathrm{f},m_\mathrm{f}}$, with $F_\mathrm{f}=3$. To model the dephasing, the absolute coupling strengths are not important; the relative strengths are computed in a straightforward manner from the Clebsch-Gordan coefficients describing the allowed transitions, once the polarizations of the control and signal fields are fixed. Over a time $t$, the atomic spins undergo Larmour precession due to the magnetic field. If the field has strength $B$ in a direction parameterized by the polar and azimuthal angles $\theta$, $\phi$, the precession is described by the operator $U = R^\dagger ER$, where $R$ rotates the quantization axis to align it with the field, and $E$ accounts for the accumulation of phase by each Zeeman level,
\begin{eqnarray}
\nonumber R &=& \sum_{m_\mathrm{i}=-F_\mathrm{i}}^{F_\mathrm{i}} e^{\mathrm{i}( Y_\mathrm{i}\sin \phi - X_\mathrm{i} \cos \phi )\theta}\ket{F_\mathrm{i},m_\mathrm{i}}\!\!\bra{F_\mathrm{i},m_\mathrm{i}} \\
\nonumber & &+ \sum_{m_\mathrm{f}=-F_\mathrm{f}}^{F_\mathrm{f}} e^{\mathrm{i}( Y_\mathrm{f}\sin \phi - X_\mathrm{f} \cos \phi )\theta}\ket{F_\mathrm{f},m_\mathrm{f}}\!\!\bra{F_\mathrm{f},m_\mathrm{f}};\\
\nonumber E &=& \sum_{m_\mathrm{i}=-F_\mathrm{i}}^{F_\mathrm{i}} e^{\mathrm{i}m_\mathrm{i}g_\mathrm{i}\mu_\mathrm{B} B t}\ket{F_\mathrm{i},m_\mathrm{i}}\!\!\bra{F_\mathrm{i},m_\mathrm{i}} \\
\nonumber & & + \sum_{m_\mathrm{f}=-F_\mathrm{f}}^{F_\mathrm{f}} e^{\mathrm{i}m_\mathrm{f}g_\mathrm{f}\mu_\mathrm{B} B t}\ket{F_\mathrm{f},m_\mathrm{f}}\!\!\bra{F_\mathrm{f},m_\mathrm{f}}.
\end{eqnarray}
Here $X_{\mathrm{i},\mathrm{f}}$, $Y_{\mathrm{i},\mathrm{f}}$ are the $x$ and $y$ components of the spin angular momentum operators for the spins $F_{\mathrm{i},\mathrm{f}}$, and $g_{\mathrm{i},\mathrm{f}}$ are the $g$-factors determining the size of the Zeeman splitting within each hyperfine manifold. In our case we have $g_\mathrm{i}=1/4$ and $g_\mathrm{f}=-1/4$. $\mu_\mathrm{B}$ is the Bohr magneton.

The Raman coherence acts as a source for the retrieved signal field in the presence of the retrieval control pulse, so the retrieval efficiency is proportional to $|\langle \Sigma \rangle |^2$, where the expectation value is evaluated just before retrieval. Incoherently summing the contributions arising from atoms starting in each of the initial Zeeman levels, we compute the retrieval efficiency according to the formula
\begin{equation}
\label{ret}
\eta(t) \propto \sum_{m_\mathrm{i}=-F_\mathrm{i}}^{F_\mathrm{i}} p_{m_\mathrm{i}} \left|\bra{F_\mathrm{i},m_\mathrm{i}} SU^\dagger \Sigma U S^\dagger \ket{F_\mathrm{i},m_\mathrm{i}}\right|^2.
\end{equation}
The only non-vanishing terms in the above expression are all proportional to $|\alpha \beta^*|^2$, confirming that the absolute values of these coefficients simply determines the scaling of $\eta$, not its shape, so that we obtain
$$
\eta(t) \propto \sum_{m_\mathrm{i}=-F_\mathrm{i}}^{F_\mathrm{i}} p_{m_\mathrm{i}} \left|\bra{F_\mathrm{i},m_\mathrm{i}} U^\dagger \Sigma U \Sigma^\dagger \ket{F_\mathrm{i},m_\mathrm{i}}\right|^2.
$$
The plot in Fig.~\ref{fig:storage} is produced by plotting $\eta(t)$, multiplied by an appropriate scaling factor, assuming a uniform initial thermal distribution $p_\mathrm{m_\mathrm{i}}=1/(2F_\mathrm{i}+1)$, where the magnetic field $B\approx 0.13$~Gauss and orientation ($\theta = 30^\circ$ from the vertical, \emph{i.e.} from the direction of the control field polarization, and $\phi = 25^\circ$ from the direction of propagation of the optical beams) are determined from a least-squares fit to the experimental data.


\end{document}